\documentclass[preprint,aps]{revtex4-1}
\usepackage{graphicx, bm, amssymb}
\usepackage[fleqn]{amsmath}
\usepackage{hyperref, cleveref}
\usepackage{setspace, float}
\usepackage{epstopdf}
\usepackage{multirow}
\begin{document}
\def\be{\begin{equation}}
\def\ee{\end{equation}}
\def\ba{\begin{array}}
\def\ea{\end{array}}
\def\bea{\begin{eqnarray}}
\def\eea{\end{eqnarray}}
\def\bc{\begin{center}}
\def\ec{\end{center}}

\title{Decay constants of pseudoscalar and vector $B$ and $D$ mesons in the light-cone quark model}
\author{Nisha Dhiman and Harleen Dahiya}
\affiliation{Department of Physics, Dr. B. R. Ambedkar National Institute of Technology, Jalandhar-144011}

\begin{abstract}
We have studied the decay constants of the pseudoscalar and the vector $B$ and $D$ mesons in the framework of the light-cone quark model. We have applied a variational method to the relativistic Hamiltonian with the Gaussian-type trial wave function to obtain the values of the scale parameter $\beta$ in different potential models. Furthermore, using the known values of the constituent quark masses of $u$, $d$, $s$, $c$ and $b$ quarks, we have obtained the numerical results for the decay constants. We have also compared our results with the other theoretical calculations and the existing experimental results. The present work predictions have many phenomenological applications in the domain of $CP$ violation and also in the determination of the CKM matrix elements.
\end{abstract}
\maketitle

\section{Introduction}
\label{sec1}
From the past few decades, a lot of theoretical and experimental efforts have been made for improving the understanding of the decay constants of heavy mesons \cite{lubicz}. Experimentally, new data on the bottom and charmed mesons decay constants ($f_B$, $f_D$ \& $f_{D_s}$) has been reported \cite{patri}.  The study of decay constants of heavy mesons with $c$ and $b$ quarks is very important, since it provides a direct source of information on the Cabibbo-Kobayashi-Maskawa (CKM) matrix elements which describe the couplings of the third generation of quarks to the lighter quarks. These matrix elements are the fundamental parameters of the Standard Model (SM) and their precise measurement will allow us to test the unitarity of the quark mixing matrix and $CP$ violation in the SM \cite{mona}. However, the uncertainty in the measurement of the decay constant makes the accurate extraction of the CKM matrix elements from the experimental data difficult. For example, in the lowest order approximation, the decay widths of the pseudoscalar and vector mesons can be written as \cite{zhi, gers}
\begin{eqnarray}
\label{eqn:0}
\Gamma (P \to \ell \nu) &=& \frac{G_F^2}{8 \pi}f_P^2 m_l^2 m_P \left(1 - \frac{m_{\ell}^2}{m_P^2}\right)^2 |V_{qQ}|^2, \quad {\rm and}\nonumber\\
\Gamma (V \to \ell \nu) &=& \frac{G_F^2}{12 \pi}f_V^2 m_V^3 \left(1 - \frac{m_{\ell}^2}{m_V^2}\right)^2 \left(1 + \frac{m_{\ell}^2}{2 m_V^2}\right) |V_{qQ}|^2. 
\end{eqnarray}
Here $G_F$ is the Fermi coupling constant, $m_{\ell}$ is the mass of the lepton, $|V_{qQ}|$ is the CKM matrix element between the constituent quarks ($q \bar{Q}$) and $m_{P(V)}$ and $f_{P(V)}$ denote the mass and the decay constant of the pseudoscalar (vector) meson, respectively. The decay constants can be regarded as the wave-function overlap of the quarks and antiquarks. The experimental measurements of the lifetimes and branching fractions of the purely leptonic decays allow us to determine the product $f_P |V_{qQ}|$ ($f_V |V_{qQ}|$). Thus, a precise theoretical input on $f_{P(V)}$ can allow a determination of the CKM matrix element.

The theoretical calculations of the decay constants of $B$ and $D$ mesons require non-perturbative treatment since at short distances, the interactions are dominated by strong force. There have been many theoretical groups that are looking on the determination of the decay constants in the realm of non-perturbative QCD using different models, such as QCD sum rules (SR) \cite{zhi, gel, narison, narison1, dominguez, baker, lucha, lucha1, mathur}, lattice QCD (LQCD) \cite{na, davies, bazavov}, relativistic quark model (RQM) \cite{capstick, colangelo, ebert, dae}. Here we focus on one such method that is useful for solving non-perturbative problems of hadron physics: the light-cone quark model (LCQM) \cite{chien, chien1, chao, ho, yang1}. In our work, we will deal with the relativistic Hamiltonian and implement the variational method using the Gaussian-type wave function in different potential models so as to obtain the ground state energy as well as the scale parameter. While performing the variational principle on the mesonic systems, the selection of the potential is very significant. H.-M. Choi \cite{ho} has obtained the decay constants by fixing the model parameters obtained from the linear and harmonic oscillator (HO) potential models within the light-front approach. The present work is devoted to the analysis of the decay constants by fixing the scale parameter under the Martin potential \cite{martin}, Cornell potential \cite{cornell}, Logarithnic potential \cite{log} and the combination of harmonic and Yukawa potentials \cite{harmonic} within the light-cone framework.

The LCQM offers many insights into the internal structures of the bound states of the mesons and thus has been widely used in the phenomenological study of the meson physics. The LCQM deals with the wave function described on the four-dimensional space-time plane defined by the equation $x^{+} = x^{0} + x^{3}$. Unlike the traditional equal-time Hamiltonian formalism, the LCQM includes the important relativistic effects in the hadronic wave functions \cite{brodsky1, wilson, dirac}. Apart from the well-known constituent masses of the quarks, the only parameter in the model is the wave function parameter $\beta$ which determines the size of the bound state and can be fixed by obtaining a fit to the data. 
%According to Dirac \cite{dirac}, there are three forms of relativistic dynamics: $(1)$ the instant form, $(2)$ the front form, and $(3)$ the point form, respectively. The front form has an advantage of the absence of square root factor in the dispersion relation
%\begin{eqnarray}
%k^- = \frac{\textbf{k}_\bot^2 + m^2}{k^+}, \nonumber
%\end{eqnarray}
%where $k^+ = k^0 + k^3$ and $k^- = k^0 - k^3$ are called light-front energy and longitudinal momentum variables, respectively and $\textbf{k}_\bot = (k^1, \ k^2)$ are orthogonal to $k^-$. This simplifies the dynamical structure of the system since the presence of the square root factor in energy-momentum relation can not provide a simple picture of the bound state Schr$\ddot{o}$edinger wave equation.
The distinguished features of the light-cone quantization approach compared to the ordinary quantization includes \cite{ho1}: $(1)$ the dynamical property of the rotation operators and $(2)$ the suppression of vacuum fluctuations. The light-cone quantization has the advantage over the ordinary equal $t$ quantization  in converting the dynamical problem from boost to rotation.
Since the rotation is compact, the rotation problem is much easier to deal with in comparison with the boost problem. The most phenomenal feature of this formalism is the simplicity of the light-cone vacuum except the zero modes. The trivial vacuum of the free light-front theory is an exact eigenstate of the total light-cone Hamiltonian\cite{brodsky2,brodsky3}.
%, i.e. there is no spontaneous creation of massive fermions in the light-cone quantized vacuum  
%The Fock space expansion constructed on this vacuum state provides a complete relativistic many-particle basis for a hadron \cite{brodsky3}. 
The light-cone wave function can be expressed in terms of  hadron momentum independent internal momentum fraction variables making it explicitly Lorentz invariant \cite{brodsky4}.

The paper is organized as follows. In Sec. \ref{sec2}, we begin with a brief description of the light-cone framework and derive the formulas for the pseudoscalar and vector mesons decay constants in LCQM. In Sec. \ref{sec3},  we first calculate the values for the scale parameter $\beta$ in different potential models using variational method with the help of Gaussian-type trial wave function and then present our numerical results for the pseudoscalar and vector $B$ and $D$ mesons, respectively. We also compare our results with the available experimental data and other theoretical model predictions. Finally, the conclusions are given in Sec. \ref{sec4}.

\section{Light-cone framework}
\label{sec2}
\subsection{General Formalism}
\label{sec:a}
We choose to work in the LCQM in which the bound state of a heavy meson composed of a light quark $q$ and a heavy antiquark $\bar{Q}$ with total momentum $P$ and spin $S$ is represented as \cite{cai}
\begin{eqnarray}
\label{eqn:1}
|M(P,S,S_z)\rangle &=& \int\frac{dp_{q}^+d^2\textbf{p}_{q_{\bot}}}{16\pi^3} \frac{dp_{\bar Q}^+d^2\textbf{p}_{\bar Q_\bot}}{16\pi^3}16\pi^3 \delta^3(\tilde P-\tilde p_{q}-\tilde p_{\bar Q}) \nonumber\\ && \times \sum\limits_{\lambda_{q},\lambda_{\bar Q}}\Psi^{SS_z}(\tilde p_{q},\tilde p_{\bar Q},\lambda_{q},\lambda_{\bar Q}) \ 
 |q(p_{q},\lambda_{q})\bar Q(p_{\bar Q},\lambda_{\bar Q})\rangle,
\end{eqnarray}
where $p_{q (\bar Q)}$ and $\lambda_{q (\bar Q)}$ are the on-mass shell light-front momentum and the light-front helicity of the constituent quark (antiquark) respectively. The four-momentum $\tilde p$ is defined as
\begin{eqnarray}
\label{eqn:2}
\tilde p=(p^+,~\textbf{p}_\perp), \ \textbf{p}_\perp=(p^1,~p^2), \ p^-=\frac{m^2+\textbf{p}_\perp^2}{p^+}, 
\end{eqnarray}
and
\begin{flalign}
\label{eqn:3}
         |q(p_{q},\lambda_{q})\bar Q(p_{\bar Q},\lambda_{\bar Q})\rangle
        &= b^\dagger(p_{q}, \lambda_{q})d^\dagger(p_{\bar Q}, \lambda_{\bar Q})|0\rangle,\nonumber \\  
        \{b(p', \lambda'),b^\dagger(p, \lambda)\} &=
        \{d(p', \lambda'),d^\dagger(p, \lambda)\} =
        2(2\pi)^3~\delta^3(\tilde p'-\tilde p)~\delta_{\lambda'\lambda}.
\end{flalign}
The light-front momenta $p_{q}$ and $p_{\bar Q}$ in terms of light-cone variables are
\begin{eqnarray}
\label{eqn:4}
p_{q}^+&=&x_1 P^+, \ \ p_{\bar Q}^+=x_2 P^+,\nonumber \\
\textbf{p}_{q_{\perp}}&=&x_1\textbf{P}_{\perp}+\textbf{k}_{\perp}, \ \ \textbf{p}_{\bar{Q}_\perp}=x_2\textbf{P}_{\perp}-\textbf{k}_{\perp},
\end{eqnarray}
where $x_{1 (2)}$ is the light-cone momentum fraction satisfying the relation $x_1 + x_2 = 1$ and $\textbf{k}_{\perp}$ is the relative transverse momentum of the constituent.\\
The momentum-space light-cone wave function $\Psi^{SS_z}$ in Eq. (\ref{eqn:1}) can be expressed as a covariant form \cite{chao, wilson, ho1, cai}
\begin{equation}
\label{eqn:5}
        \Psi^{SS_z}(\tilde p_{q},\tilde p_{\bar Q},\lambda_{q},\lambda_{\bar Q})
                ={\sqrt{p_{q}^+p_{\bar Q}^+}\over \sqrt{2} ~{\sqrt{{M_0^2} - (m_{q} - m_{\bar Q})^2}}}
        ~\bar u(p_{q},\lambda_{q})\,\Gamma\, v(p_{\bar Q},\lambda_{\bar Q})~\sqrt{dk_z\over dx}~\phi(x, \textbf{k}_\bot),
\end{equation}
where $\phi(x, \textbf{k}_\bot)$ describes the momentum distribution of the constituents in the bound state with $x \equiv x_2$ and
\begin{eqnarray}
\label{eqn:6}
M^2_0&=& \frac{m^2_{q}+\textbf{k}_\bot^2}{x_1}
         + \frac{m^2_{\bar Q}+\textbf{k}_\bot^2}{x_2},
\end{eqnarray}
$M_0$ defines the invariant mass of the quark system $q \bar{Q}$ which is generally different from the mass $M$ of the meson because the meson, quark and antiquark cannot be simultaneously on shell.
Also,
\begin{eqnarray}
\label{eqn:7}
&& {\rm for \, \, \, pseudoscalar \, \, \, meson},  \,\,\,\,      \Gamma=\gamma_5  , \,{\rm and} \nonumber \\
&& {\rm for \, \, \, vector \, \, \, meson},  \,\,\,\,       \Gamma=-\not{\! \hat{\varepsilon}}(S_z)+{\hat{\varepsilon}\cdot(p_{q}-p_{\bar Q})
                \over M_0+m_{q}+m_{\bar Q}},
\end{eqnarray}
with
\begin{eqnarray}
\label{eqn:8}
        &&	\hat{\varepsilon}^\mu(\pm 1) =
                \left[{2\over P^+} \vec \varepsilon_\bot (\pm 1) \cdot
                \vec P_\bot,\,0,\,\vec \varepsilon_\bot (\pm 1)\right], \nonumber\\
            &&   \vec \varepsilon_\bot
                (\pm 1)=\mp(1,\pm i)/\sqrt{2} , \nonumber\\
        &&\hat{\varepsilon}^\mu(0)={1\over M_0}\left({-M_0^2+P_\bot^2\over
                P^+},P^+,P_\bot\right).
\end{eqnarray}
The Dirac spinors satisfy the relations
\begin{eqnarray}
\label{eqn:9}
&&\sum\limits_{\lambda}u(p,\lambda)\bar u(p,\lambda)=\frac{
  \rlap{\hspace{0.03cm}/}{p}+m}{p^+} \,\,\, {\rm for ~ quark ~ and} ,\nonumber\\
 &&\sum\limits_{\lambda}v(p,\lambda)\bar v(p,\lambda)=
  \frac{\rlap{\hspace{0.03cm}/}{p}-m}{p^+} \,\,\, {\rm for ~ antiquark} .
\end{eqnarray}
The meson state can be normalized as
\begin{eqnarray}
\label{eqn:10}
        \langle M(P',S',S'_z)|M(P,S,S_z)\rangle = 2(2\pi)^3 P^+
        \delta^3(\tilde P'- \tilde P)\delta_{S'S}\delta_{S'_zS_z}~,
\end{eqnarray}
in order that
\begin{eqnarray}
\label{eqn:11}
        \int {dx\,d^2\textbf{k}_\bot}~{dk_z\over dx}~|\phi(x,\textbf{k}_\bot)|^2 = 1. 
\end{eqnarray}
We use the Gaussian-type wave function to describe the radial wave function $\phi$
\begin{eqnarray}
\label{eqn:12}
\phi(x,\textbf{k}_\bot)=\frac{1}{(\sqrt{\pi}\beta)^{3/2}}
\exp(-\textbf{k}^{2}/2\beta^{2}),
\end{eqnarray}
where $\beta$ represents the scale parameter and $\textbf{k}^2=\textbf{k}^2_\bot + k^2_z$ is the internal momentum of the meson. The longitudinal component $k_z$ is defined by
\begin{eqnarray}
\label{eqn:13}
k_z &=& (x-\frac{1}{2})M_0 + \frac{m^2_{q}-m^2_{\bar Q}}{2M_0},\, \, {\rm and}\nonumber \\
 {dk_z\over dx} &=& \frac{M_{0}}{4x(1-x)}
\biggl[1-\biggl(\frac{m^2_q-m^2_{\bar{Q}}}{M^2_{0}}\biggr)^2
\biggr].
\end{eqnarray}

\subsection{Decay constants of the pseudoscalar and the vector mesons}
\label{sec:b}
The pseudoscalar and the vector meson decay constants are defined through the matrix elements of axial and vector currents between the meson state and the vacuum \cite{ho}, i.e.
\begin{eqnarray}
\label{eqn:14}
\langle 0|A^\mu|P(P) \rangle &=& if_P P^\mu, \nonumber\\
\langle 0|V^\mu|V (P) \rangle &=& f_V M_V \epsilon^{\mu}.
\end{eqnarray}
Here $P$ is the meson momentum, $M_V$ is the mass of the vector meson, and $\varepsilon^\mu$ is the polarization vector, respectively.
These matrix elements can be solved using the formalism described in Sec. \ref{sec:a}:
\begin{eqnarray}
\label{eqn:15}
\langle 0|A^\mu|P (P) \rangle = \langle 0|\bar{Q} \gamma^\mu \gamma_5 q|P (P) \rangle =&& \int\frac{dp_{q}^+d^2\textbf{p}_{q_{\bot}}}{16\pi^3} \frac{dp_{\bar Q}^+d^2\textbf{p}_{\bar Q_\bot}}{16\pi^3}16\pi^3 \delta^3(\tilde P-\tilde p_{q}-\tilde p_{\bar Q}) \nonumber\\ 
&& \times \sum\limits_{\lambda_{q},\lambda_{\bar Q}}\Psi^{0 0} \times \langle 0|\bar{Q} \gamma^\mu \gamma_5 q|q \bar Q\rangle, \ {\rm and} \nonumber\\
\langle 0|V^\mu|V (P) \rangle = \langle 0|\bar{Q} \gamma^\mu  q|V (P) \rangle =&& \int\frac{dp_{q}^+d^2\textbf{p}_{q_{\bot}}}{16\pi^3} \frac{dp_{\bar Q}^+d^2\textbf{p}_{\bar Q_\bot}}{16\pi^3}16\pi^3 \delta^3(\tilde P-\tilde p_{q}-\tilde p_{\bar Q}) \nonumber\\ 
&& \times \sum\limits_{\lambda_{q},\lambda_{\bar Q}}\Psi^{1 S_z} \times \langle 0|\bar{Q} \gamma^\mu q|q \bar Q\rangle.
\end{eqnarray}
Using the light-cone wave function, the decay constants of the pseudoscalar and vector mesons are given by
\begin{eqnarray}
\label{eqn:16}
f_P = 2\sqrt{6}\int dxd^2\textbf{k}_{\perp} \sqrt{dk_z\over dx}\phi(x,\textbf{k}_\perp) \frac{\mathcal{A}}{\sqrt{\mathcal{A}^2+\textbf{k}^2_{\perp}}},
\end{eqnarray}
\begin{eqnarray}
\label{eqn:17}
f_V &=& 2\sqrt{6}\int dxd^2\textbf{k}_\perp \sqrt{dk_z\over dx}\frac{\phi(x,\textbf{k}_\perp)}{\sqrt {{\cal A}^2+\textbf{k}_\perp^2}}\frac{1}{M_{0}}
\Big\{m_q m_{\bar Q}+x(1-x)M_{0}^2+\nonumber \\
&&\textbf{k}_\perp^2 +\frac{{\cal B}}{2W}\left[\frac{m_q^2+\textbf{k}_\perp^2}{1-x}-\frac{m_{\bar Q}^2+\textbf{k}_\perp^2}{x}-(1-2x)M_{0}^2\right]\Big\}.
\end{eqnarray}
where
\begin{eqnarray}
&&\mathcal{A} = m_q x + m_{\bar Q} (1 - x), \nonumber\\
&&\mathcal{B} = m_q x - m_{\bar Q} (1 - x), \nonumber\\
&&W = M_0 + m_q +m_{\bar Q}. \nonumber
\end{eqnarray}
For a given value of $\beta$, $f_P$ and $f_V$ can be calculated from Eqs. (\ref{eqn:16}) and (\ref{eqn:17}) using the constituent quark masses of $u$, $d$, $s$, $c$ and $b$ quarks, respectively.

\section{Numerical Results}
\label{sec3}
\begin{figure}[h]
\centering
\minipage{0.5\textwidth}
  \includegraphics[width=3 in]{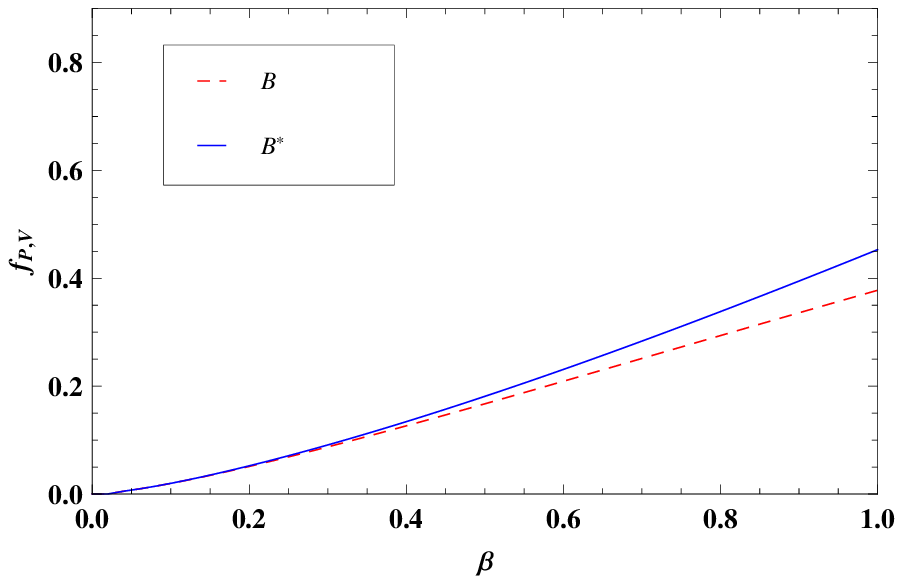}
\endminipage\hfill
\minipage{0.5\textwidth}
 \includegraphics[width=3 in]{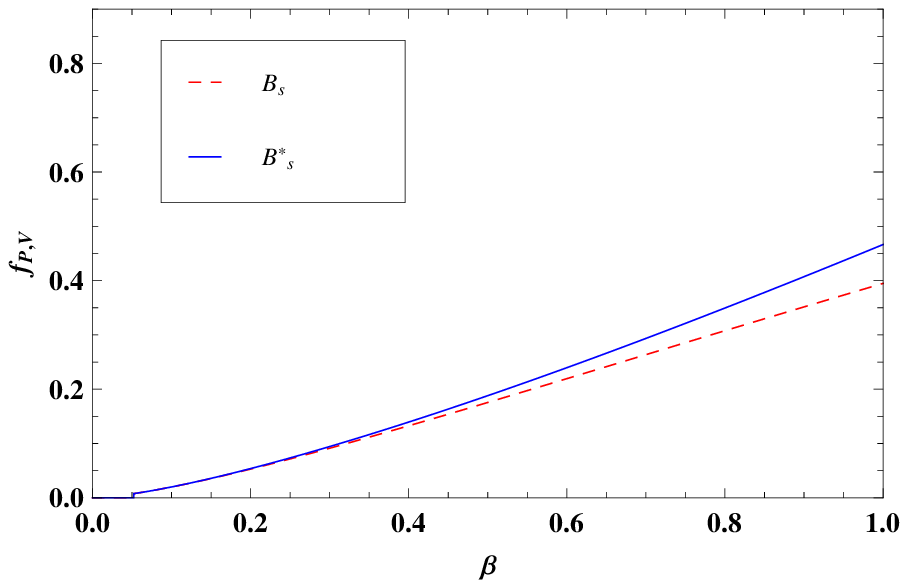}
\endminipage\hfill
\caption{The pseudoscalar and the vector decay constants for $B$ and $B_s$ mesons ($f_{B}$ and $f_{B^*}$ in the left panel, $f_{B_s}$ and $f_{B^*_s}$ in the right panel) as functions of the parameter $\beta$ (in GeV).} 
\label{fig:1}
\end{figure}

\begin{figure}[h]
\centering
\minipage{0.5\textwidth}
  \includegraphics[width=3 in]{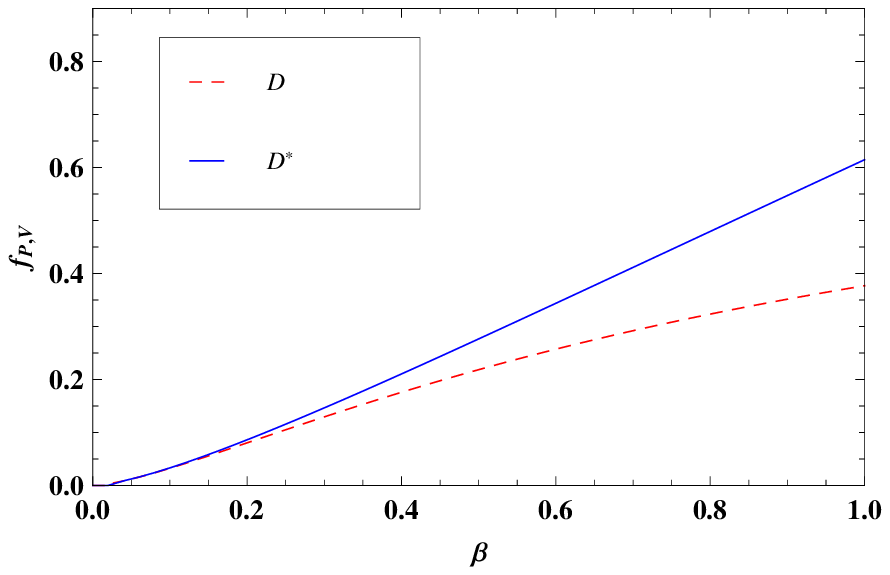}
\endminipage\hfill
\minipage{0.5\textwidth}
 \includegraphics[width=3 in]{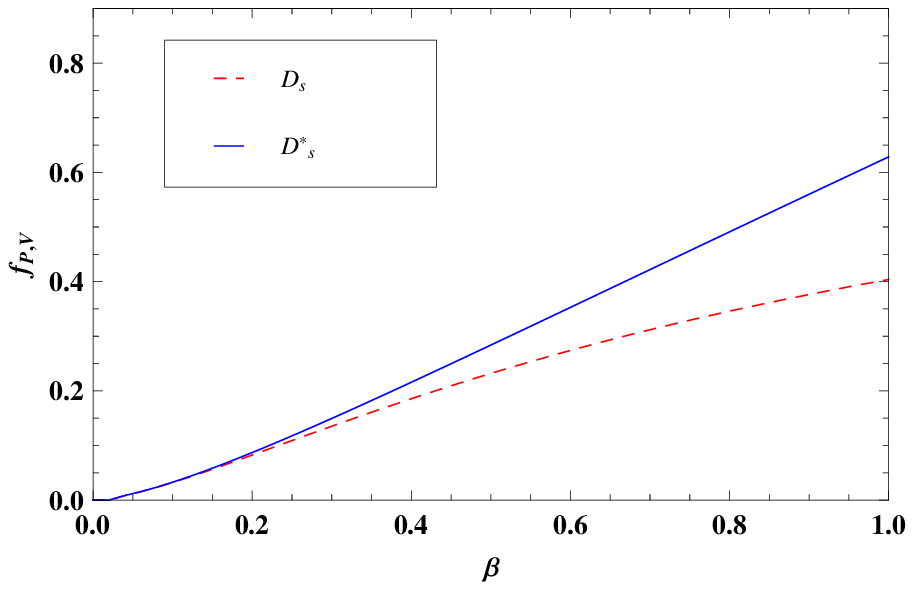}
\endminipage\hfill
\caption{The pseudoscalar and the vector decay constants for $D$ and $D_s$ mesons ($f_{D}$ and $f_{D^*}$ in the left panel, $f_{D_s}$ and $f_{D^*_s}$ in the right panel) as functions of the parameter $\beta$ (in GeV).} 
\label{fig:2}
\end{figure}

\begin{figure}[h]
  \begin{center}
    \includegraphics[width=3.5 in]{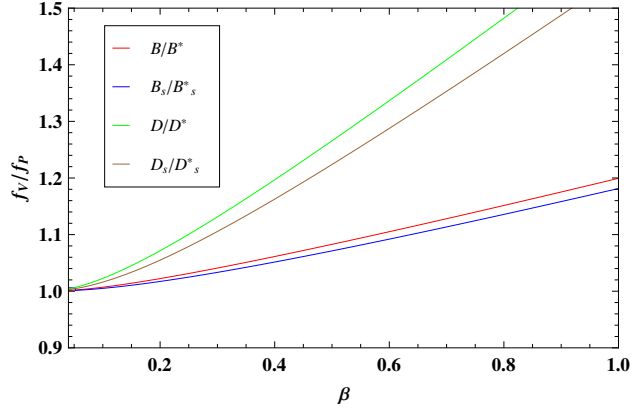}
  \end{center}
  \caption{The ratios of pseudoscalar and the vector decay constants for the  $B$, $B_s$, $D$ and $D_s$ mesons ($f_{B^*}/f_{B}$, $f_{B^*_s}/f_{B_s}$, $f_{D^*}/f_{D}$ and $f_{D^*_s}/f_{D_s}$) as functions of the parameter $\beta$ (in GeV).} 
  \label{fig:3}
\end{figure}

Numerically, we obtain the pseudoscalar and the vector meson decay constants for $B$, $B_s$, $D$ and $D_s$ mesons as functions of the scale parameter $\beta$, using the following values of constituent quark masses
\begin{center}
$m_u = m_d = 0.25$ GeV, $m_s = 0.38$ GeV, $m_c = 1.5$ GeV and $m_b = 4.8$ GeV.
\end{center}
These quark masses have been fixed using the experimentally known values of the decay constants \cite{chao}.
In order to understand the dependence of the decay constants on the scale parameter $\beta$, we present the decay constants $f_P$ and $f_V$ as functions of $\beta$ for $B$, $B_s$, $D$ and $D_s$ mesons in Figs. \ref{fig:1} and \ref{fig:2}, respectively.
Even though it is expected that the pseudoscalar and the vector meson decay constants will have similar values because they differ with each other only in their internal spin configurations as is clear from Eqs. (\ref{eqn:16}) and (\ref{eqn:17}), however, to exploit their quantitative difference, we present in Fig. \ref{fig:3} our results for the ratios $f_V/ f_P$ as a function of $\beta$. It is observed from Figs. \ref{fig:1}$-$\ref{fig:3} that the decay constants increase as the value of the parameter $\beta$ increases and the difference between the vector and pseudoscalar decay constant also increases with the increasing $\beta$ values. It is clear from Figs. \ref{fig:1} and \ref{fig:2} that the vector decay constants are higher as compared to the pseudoscalar decay constants at higher values of $\beta$. This is true for the cases of all the mesons $B$, $B_s$, $D$ and $D_s$. The rise is however steeper in the case of $D$ and $D_s$ mesons when compared with the $B$ and $B_s$ mesons. This is clearly evident in Fig. \ref{fig:3} where the ratios $f_V/ f_P$ for the cases of $D$ and $D_s$ mesons are almost 1.5 times higher than the $B$ and $B_s$ mesons. This variation of decays constants with $\beta$ clearly indicates that in order to get the reliable results for the decay constants, it is essential to have accurate values of $\beta$.

In the heavy ($B$ and $D$) mesons systems, we treat the motion of heavy ($b$ and $c$) quarks non-relativistically, but the motion of light quarks is treated relativistically. So in the present work, we apply the variational method to the following relativistic Hamiltonian to obtain the precise values of $\beta$ \cite{hwang}:
\begin{eqnarray}
\label{eqn:18}
H = \sqrt{\textbf{k}^{2} + m^2_{q}} + \sqrt{\textbf{k}^{2} + m^2_{\bar Q}} + V (r),
\end{eqnarray}
where $\textbf{k} = (\textbf{k}_\bot, k_z)$ is the three-momentum of the constituent quark. In the variational method, the expectation value of the Hamiltonian is calculated with some trial wave function consisting of a variational parameter whose value is determined by the stationary condition. Here we consider the Gaussian wave function in Eq. (\ref{eqn:11}) as our trial wave function with the variational parameter $\beta$. The Fourier transform of $\phi (\textbf{k})$ gives us the coordinate space wave function $\psi (\textbf{r})$, which is also Gaussian
\begin{eqnarray}
\label{eqn:19}
\psi(\textbf{r})=\left(\frac{\beta}{\sqrt{\pi}}\right)^{3/2}
\exp\left(-\beta^2 \textbf{r}^{2}/2\right).
\end{eqnarray}
We can obtain the ground state energy for our system by minimizing the expectation value of the Hamiltonian $H$, $\langle H\rangle = \langle\psi ({\bf r})|H|\psi ({\bf r})\rangle = \langle\phi ({\bf k})|H|\phi ({\bf k})\rangle = E (\beta)$, that is,
\begin{eqnarray}
\label{eqn:19a}
\frac{d E (\beta)}{d \beta} = 0 \quad {\rm at}\,\, \beta = \bar{\beta}, \nonumber
\end{eqnarray}
where $\bar{\beta}$ denotes the inverse size of the meson ($\langle r^2 \rangle^{1/2} = 3/(2 \bar{\beta})$) and $E(\bar \beta) = \bar E$ approximates the meson mass $m_M$. Various potential models have been used in the literature to obtain the accurate values of the parameter $\beta$, however, in the present work, we have considered the potential $V (r)$ in Eq. (\ref{eqn:18}) from the different potential models listed in Table \ref{tab1}.
 
\begin{table}[h]
\centering
\caption{Form of the potential $V(r)$ in different models and their respective parameters.}
\vspace{0.2cm}
\label{tab1}
\begin{tabular}{ll}
\hline\hline\\[-1.5 ex]
\multicolumn{1}{c}{Potential Model}                                                                                  & \multicolumn{1}{c}{Form of the Potential} \\[1.5 ex] \hline\\[-1.5 ex]
Martin Potential \cite{martin}                                                                                           & \hspace{0.25in} $V (r) = -8.064 + 6.898\,\, r^{0.1}$.                                           \\[2 ex]
\begin{tabular}[c]{@{}l@{}}Cornell Potential of\\ Hagiwara \textit{et al.} \cite{cornell} \end{tabular}                                       & \hspace{0.3in}\begin{tabular}[c]{@{}l@{}}$V(r) = \frac{- \alpha_c}{r} + K r$,\\ with $\alpha_c = 0.47$ and $K = 0.19$ {GeV}${^2}$.\end{tabular}                                        \\[4 ex]                                          
Logarithmic Potential \cite{log}                                                                                                & \hspace{0.25in} $V(r) = - 0.6635 + 0.733\,\, {\rm ln}\,\, r$.                                          \\[2 ex]                                          
\begin{tabular}[c]{@{}l@{}}Combination of harmonic \\ and Yukawa potentials \cite{harmonic}     \end{tabular} & \hspace{0.3in} \begin{tabular}[c]{@{}l@{}}$V (r) = a r^2 + b \frac{e^{-\alpha r}}{r}$,\\ with $b = - 0.42$ and $\alpha = 0.01$ GeV.\end{tabular}                                                                                  \\[4 ex]                                           \hline\hline
\end{tabular}
\end{table}

\begin{figure}[h]
  \begin{center}
\includegraphics[width=4.5 in]{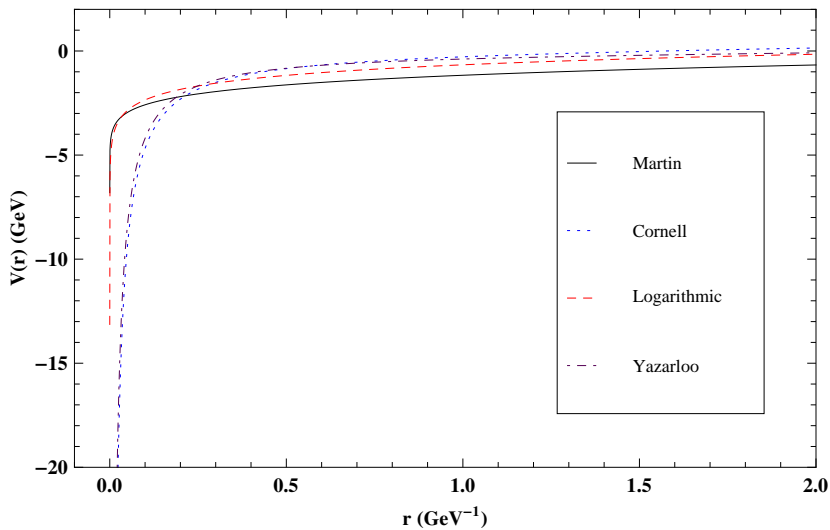}
  \end{center}
  \caption{Variation of $V(r)$ with respect to $r$ in various potential models.} 
  \label{fig:4}
\end{figure}

For the sake of comparison, we have also shown the plots of $V (r)$ versus $r$ for all the potential models that we have considered in this work in Fig. \ref{fig:4}. We note that the different potentials are pretty similar in the relevant range of $0.3 \leq r \leq 2.0$ GeV$^{-1}$ to each other.

With the help of the Gaussian trial wave function in Eq. (\ref{eqn:12}) (or (\ref{eqn:19})), we can now calculate the expectation value of the Hamiltonian (\ref{eqn:18}).
Our results for the variational parameter $\beta$ of the Gaussian wave function in different potential models has been listed in Table \ref{tab2}. 
From Table \ref{tab2}, we notice that the $\bar{\beta}$ values follow a similar pattern, that is, $\bar{\beta}_{B_s} > \bar{\beta}_{B} > \bar{\beta}_{D_s} > \bar{\beta}_{D}$ in each of the potential model. As the masses of the quarks inside the meson increase, the distance between them decreases making $\bar{\beta}$ increase with the increase in the quark masses. It is important to mention here that $\bar{\beta}$  represents the inverse size of the meson. Also, Table \ref{tab2} gives us the following average values of $\bar{E}$ which approximate the meson masses:
\begin{eqnarray}
\label{eqn:20}
\bar{E}_{B} = 5.26 \, {\rm GeV}, \quad \bar{E}_{B_s} = 5.34 \, {\rm GeV}, \nonumber\\
\bar{E}_{D} = 2.04 \, {\rm GeV}, \quad \bar{E}_{D_s} = 2.14 \, {\rm GeV}.
\end{eqnarray}

\begin{table}[h]
\centering
\caption{Variational parameter $\beta$ and the corresponding values of minimum energy $E (\bar{\beta})$ (in units of GeV) for Gaussian-type wave function in different potential models.}
\vspace{0.2cm}
\label{tab2}
\begin{tabular}{lclclclcl}
\hline\hline\\[-1.5 ex]
\multicolumn{1}{c}{Potential Model}                                                     & $\bar{\beta}_{B}$     & \multicolumn{1}{c}{$\bar{E}_{B}$} & $\bar{\beta}_{B_s}$     & \multicolumn{1}{c}{$\bar{E}_{B_s}$} & $\bar{\beta}_{D}$     & \multicolumn{1}{c}{$\bar{E}_{D}$} & $\bar{\beta}_{D_s}$     & \multicolumn{1}{c}{$\bar{E}_{D_s}$}
\\[1 ex]
\hline\\[-1.5 ex]
Martin \cite{martin}                                                                                  &\hspace{0.1in} 0.592 &\hspace{0.01in}    4.803                   &\hspace{0.01in}    0.628 &\hspace{0.01in}                     4.869     &\hspace{0.01in}    0.494 &\hspace{0.01in} 1.603                      &\hspace{0.01in}    0.528 &\hspace{0.01in}    1.682                      \\[1 ex]
Cornell \cite{cornell}                                                                                 &\hspace{0.1in} 0.561 &\hspace{0.01in}    5.624                      &\hspace{0.01in}    0.600 &\hspace{0.01in}    5.692                      &\hspace{0.01in}    0.476 &\hspace{0.01in}    2.414                      &\hspace{0.01in}    0.511 &\hspace{0.01in}      2.496                    \\[1 ex]
Logarithmic \cite{log}                                                                             &\hspace{0.1in} 0.595 &\hspace{0.01in}     5.311                     &\hspace{0.01in}    0.633 &\hspace{0.01in}      5.376                    &\hspace{0.01in}    0.489 &\hspace{0.01in}    2.110                      &\hspace{0.01in}    0.527 &\hspace{0.01in}      2.189                    \\[1 ex]
\begin{tabular}[c]{@{}l@{}}Combination of Harmonic\\ and Yukawa Potentials \cite{harmonic}                    \end{tabular} &\hspace{0.1in} 0.408 &\hspace{0.01in}    5.297                      &\hspace{0.01in}    0.496 &\hspace{0.01in}    5.436                      &\hspace{0.01in}    0.362 &\hspace{0.01in}    2.034                      &\hspace{0.01in}    0.436 &\hspace{0.01in}    2.190                     \\[2 ex]
\hline\hline
\end{tabular}
\end{table}

The calculated values of masses for $B$ meson are in good agreement when compared to the measured values whereas the results for $D$ mesons are on the slightly larger side \cite{patri} ($m_B = 5.28$ GeV, $m_{B_s} = 5.37$ GeV, $m_D = 1.87$ GeV and $m_{D_s} = 1.97$ GeV ). In the present work we have not included the spin-dependent interaction term in the Hamiltonian (\ref{eqn:18})  considering it as a perturbation term. Including this term however introduces the spin-corrections to the wave function and we get reasonable values for the masses. Using the values of $\beta$ from Table \ref{tab2}, we can calculate the decay constants for $B$ and $D$ mesons by using the Eqs. (\ref{eqn:16}) and (\ref{eqn:17}), respectively. The numerical results of the decay constants that we obtained have been listed in Table \ref{tab3}.

\begin{table}[h]
\centering
\caption{Pseudoscalar and vector meson decay constants for $B$ and $D$ mesons (in units of MeV) in different potential models.}
\vspace{0.2cm}
\label{tab3}
\begin{tabular}{lclclclcl}
\hline\hline\\[-1.5 ex]
\multicolumn{1}{c}{Potential Model} \hspace{1.0in}                                                     &\multicolumn{1}{l} {$f_{B}$} \hspace{0.1in}     & \multicolumn{1}{l}{$f_{B^*}$} \hspace{0.1in} &\multicolumn{1}{l} {$f_{B_s}$} \hspace{0.1in}     & \multicolumn{1}{l}{$f_{B_s^*}$} \hspace{0.1in} &\multicolumn{1}{l} {$f_{D}$} \hspace{0.1in}     & \multicolumn{1}{l}{$f_{D^*}$} \hspace{0.1in} &\multicolumn{1}{l} {$f_{D_s}$} \hspace{0.1in}     & \multicolumn{1}{l}{$f_{D_s^*}$}
\\[1 ex]
\hline\\[-1.5 ex]
Martin \cite{martin}                                                                                  &\multicolumn{1}{l} {206} &\multicolumn{1}{l} {227}                   &\multicolumn{1}{l}  {232} &\multicolumn{1}{l} {254}     &\multicolumn{1}{l}    {216} &\multicolumn{1}{l} {272}                      &\multicolumn{1}{l}    {244} &\multicolumn{1}{l} {303}                      \\[1 ex]
Cornell \cite{cornell}                                                                                 &\multicolumn{1}{l} {193} &\multicolumn{1}{l} {211}                      &\multicolumn{1}{l} {219} &\multicolumn{1}{l} {239}                      &\multicolumn{1}{l} {209} &\multicolumn{1}{l} {260} &\multicolumn{1}{l}  {237} &\multicolumn{1}{l} {291}                    \\[1 ex]
Logarithmic \cite{log}                                                                             &\multicolumn{1}{l} {207} &\multicolumn{1}{l} {228}                     &\multicolumn{1}{l} {234} &\multicolumn{1}{l} {257} &\multicolumn{1}{l}  {214} &\multicolumn{1}{l} {269}                      &\multicolumn{1}{l} {244} &\multicolumn{1}{l} {302}                    \\[1 ex]
\begin{tabular}[c]{@{}l@{}} Combination of Harmonic\\ and Yukawa Potentials \cite{harmonic} \end{tabular} &\multicolumn{1}{l} {129} &\multicolumn{1}{l}    {138}                      &\multicolumn{1}{l} {174} &\multicolumn{1}{l}   {186}                      &\multicolumn{1}{l} {159} &\multicolumn{1}{l}   {186} &\multicolumn{1}{l}  {203} &\multicolumn{1}{l}   {240} 
\\[2 ex]
%\\[4.5 ex]
%Average&\multicolumn{1}{c} {$184 \pm 32$} &\multicolumn{1}{c}     {$201 \pm 37$}                     &\multicolumn{1}{c}    {$215 \pm 24$} &\multicolumn{1}{c}      {$234 \pm 29$} &\multicolumn{1}{c}    {$200 \pm 24$} &\multicolumn{1}{c}   {$247 \pm 35$} &\multicolumn{1}{c}    {$232 \pm 17$} &\multicolumn{1}{c}     {$284 \pm 29$}       \\[1 ex]
\hline\hline
\end{tabular}
\end{table}

After obtaining the decay constants, we can obtain the ratios for the vector and the pseudoscalar $B$ and $D$ mesons decay constants. The ratios $f_{B^*}/f_{B}$, $f_{B_{s^*}}/f_{B_s}$, $f_{D^*}/f_{D}$ and $f_{D_{s^*}}/f_{D_s}$ in different potential models have been listed in Table \ref{tab45}. It is observed that the ratios $f_V/f_P$ for $D$ mesons are larger as compared to $B$ mesons which can be understood by looking into the last term appearing in Eq. (\ref{eqn:17}).  The mean square value of the transverse momentum $\bf{k_\perp}$ in the Gaussian wave function is directly proportional to the square of  $\beta$ ($<\textbf{k}^2_\perp>\, \propto \beta^2$), indicating that the value of $\beta$ affects the ratio  $f_V/f_P$. It is also important to mention here that the ratios $f_V/f_P$ of the vector and the pseudoscalar heavy mesons depend on the heavy quark masses. 
%On the one hand, the dependence of the ratios $f_V/f_P$ on the $b$ and $c$ masses  varies as  $m_b >> m_c$ whereas on the other hand the scale parameter for $B$ and $D$ mesons has opposite effect with $\beta_B > \beta_D$. 
Therefore,  even though $m_b >> m_c$ and $\beta_B > \beta_D$ but because of the large difference between the $b$ and $c$ masses,  the quark mass effect dominates over the scale parameter leading to the ratio $f_V/f_P$ for $D$ mesons being larger than that of $B$ mesons. These results are also in agreement with the model-independent analysis
of semileptonic $B$ meson decays in the context of the heavy quark
effective theory \cite{hqet-neubert}.

\begin{table}[h]
\centering
\caption{Ratios $f_{B^*}/f_{B}$, $f_{B_{s^*}}/f_{B_s}$, $f_{D^*}/f_{D}$ and $f_{D_{s^*}}/f_{D_s}$ in different potential models.}
\vspace{0.2cm}
\label{tab45}
\begin{tabular}{lcccc}
\hline\hline\\[-1.5 ex]
\multicolumn{1}{c}{Potential Model} \hspace{1.0in}  &  \multicolumn{1}{l} {$f_{B^*}/f_{B}$} \hspace{0.2in}   & \multicolumn{1}{l} {$f_{B_{s^*}}/f_{B_s}$} \hspace{0.2in}  & \multicolumn{1}{l} {$f_{D^*}/f_{D}$} \hspace{0.2in}     & \multicolumn{1}{l} {$f_{D_{s^*}}/f_{D_s}$} 
\\[1 ex]
\hline\\[-1.5 ex]
Martin \cite{martin} 
& \multicolumn{1}{l} {1.10} & \multicolumn{1}{l} {1.09} & \multicolumn{1}{l} {1.26} & \multicolumn{1}{l} {1.24}  \\[1 ex]
Cornell \cite{cornell}                                                                                 & \multicolumn{1}{l} {1.09} & \multicolumn{1}{l} {1.09} & \multicolumn{1}{l} {1.24} & \multicolumn{1}{l} {1.23}  \\[1 ex]
Logarithmic \cite{log}                                                                             & \multicolumn{1}{l} {1.10} & \multicolumn{1}{l} {1.10} & \multicolumn{1}{l} {1.26} & \multicolumn{1}{l} {1.24}  \\[1 ex]
\begin{tabular}[c]{@{}l@{}}Combination of Harmonic\\ and Yukawa Potentials \cite{harmonic}                    \end{tabular} & \multicolumn{1}{l} {1.07} & \multicolumn{1}{l} {1.07} & \multicolumn{1}{l} {1.17} & \multicolumn{1}{l} {1.18}  \\[2 ex]
\hline\hline
\end{tabular}
\end{table}

To compare our predictions for the bottom and charmed mesons decay constants with the available experimental data \cite{patri} and the other theoretical calculations, we have presented in Tables \ref{tab4} and \ref{tab5} our results. The results from  the QCD SR \cite{zhi, gel, narison}, LQCD \cite{na, davies}, RQM \cite{capstick, ebert, colangelo, dae}, Bethe-Salpeter (BS) \cite{cevi, guo} and LFQM \cite{chien, chao, ho}  have also been presented in the tables. The existing experimental results for the decay constants are available only for $f_B$, $f_D$ and $f_{D_s}$ \cite{patri}. We note that our predictions for the decay constants of $B$ and $D$ mesons are more or less in agreement with the available experimental data. Also, we note that the theoretical results predicted by various models (including this work) differ from each other in one way or the other. We can see from Table \ref{tab4} that our results for the decay constants of $B$ mesons under the Martin \cite{martin}, Cornell \cite{cornell} and logarithmic \cite{log} potentials are in good agreement with the QCD SR results \cite{zhi, gel, narison}. Similarly, from Table \ref{tab5} we find that our results for the decay constants of $D$ mesons are consistent with the ones obtained from the QCD SR \cite{gel} as well as from the linear \{HO\} parameters \cite{ho}. The difference in the values of the decay constants with respect to other theoretical models might be due to different model assumptions or distinct choices of the parameters. However, overall the results are very much in the same range. The present predictions for the pseudoscalar and the vector mesons decay constants are important and  have many phenomenological implications especially in studying the $CP$ violation and in extracting the CKM matrix elements. 

\section{Conclusions}
\label{sec4}
In this work, we have studied the decay constants of the pseudoscalar and the vector $B$ and $D$ mesons within the framework of the LCQM. We have calculated the values of the scale parameter $\beta$ in different potential models from the variational calculation of the relativistic Hamiltonian using the Gaussian-type trial wave function. Using the known values of the constituent quark masses of $u$, $d$, $s$, $c$ and $b$ quarks and the calculated values of the parameter $\beta$, we have obtained the decay constants of the pseudoscalar and the vector $B$ and $D$ mesons in different potential models, respectively. Our predictions for the decay constants in the LCQM are more or less in agreement with the available experimental results as well as the other existing theoretical model predictions. The future experiments to measure the  decay constants $f_B^*$, $f_{B_s}$, $f_{B_s}^*$, $f_D^*$ and $f_{D_s}^*$ will not only provide a direct way to determine the decay constants and the scale parameter but will also impose significant constraint on CP violation and in extracting the CKM matrix elements precisely.

\section*{Acknowledgements}
Authors would like to thank the Department of Science and Technology (Ref No. SB/S2/HEP-004/2013) Government of India for financial support.

\begin{table}[H]
	\centering
	\caption{Pseudoscalar and vector $B$ mesons decay constants (in units of MeV) in the present work and their comparison with experimental and other theoretical model predictions.}
	\vspace{0.2 cm}
	\label{tab4}
	\begin{tabular}{l p{1.5 in} cccc} 
	 \hline\hline\\[-2.5 ex]
		{} & {} & {$f_{B}$} & {$f_{B^*}$} & {$f_{B_s}$} & {$f_{B_s^*}$} \\[1 ex]
		\hline\\[-1.3 ex]
		\multirow{4}{8em}{Present work with different potential models} & {Martin} & {206} & {227} & {232} & {254} \\
		& {Cornell} & {193} & {211} & {219} & {239} \\                   
		& {Logarithmic} & {207} & {228} & {234} & {257} \\
		& {Harmonic plus Yukawa} & {129} & {138} & {174} & {186} \\[1 ex]
		\hline\\[-3.5 ex]
		\multirow{1}{5em}{} & Experimental \cite{patri} &  {$188(17)(18)$} & {$-$} & {$-$} & {$-$} \\[1 ex]
		 \hline\\[-1.3 ex]
		 \multirow{11}{10em} {Other theoretical predictions} & QCD SR \cite{zhi} & {$194 \pm 15$} & {$213 \pm 18$} & {$231 \pm 16$} & {$255 \pm 19$} \\[1ex]
		& QCD SR \cite{gel} & {$207^{+17}_{-9}$} & {$210^{+10}_{-12}$} & {$242^{+17}_{-12}$} &   {$251^{+13}_{-16}$} \\[1ex]
		& QCD SR \cite{narison} & {$206 \pm 7$} & {$-$} & {$234 \pm 5$} & {$-$} \\[1ex]
		& LQCD \cite{na} & {$191 \pm 9$} & {$-$} & {$228 \pm 10$} & {$-$} \\[1ex]
		& RQM \cite{capstick} & {$155 \pm 15$} & {$-$} & {$210 \pm 20$} & {$-$} \\[1ex]
		& RQM \cite{ebert} & {$189$} & {$219$} & {$218$} & {$251$} \\[1ex] 
		& RQM \cite{dae} & {$231 \pm 9$} & {$252 \pm 10$} & {$266 \pm 10$} & {$289 \pm 11$} \\[1ex]
		& BS \cite{cevi, guo} & {$196 \pm 29$} & {$238 \pm 18$} & {$216 \pm 32$} & {$272 \pm 20$} \\[1ex]
		& LFQM \cite{chien} &  {$-$} & {$225 \pm 38$} & {$281 \pm 54$} & {$313 \pm 67$} \\[1ex]
		& LFQM \cite{chao} &  {$-$} & {$201.9^{+43.2}_{-41.4}$} & {$-$} & {$244.2 \pm 7.0$} \\[1ex] 
		& Linear\{HO\} \cite{ho} &  {$189 \{180\}$} & {$204 \{193\}$} & {$234 \{237\}$} & {$250 \{254\}$} \\[1.5 ex]
		\hline\hline
	\end{tabular}
\end{table}

\begin{table}[H]
	\centering
	\caption{Pseudoscalar and vector $D$ mesons decay constants (in units of MeV) in the present work and their comparison with experimental and other theoretical model predictions.}
	\vspace{0.2 cm}
	\label{tab5}
	\begin{tabular}{l p{1.5 in} cccc} 
	 \hline\hline\\[-2.5 ex]
	 {} & {} & {$f_{D}$} & {$f_{D^*}$} & {$f_{D_s}$} & {$f_{D_s^*}$} \\[1 ex]
	 \hline\\[-1.3 ex]
		\multirow{4}{8em}{Present work with different potential models} & {Martin} & {216} & {272} & {244} & {303} \\
		& {Cornell} & {209} & {260} & {237} & {291} \\                   
		& {Logarithmic} & {214} & {269} & {244} & {302} \\
		& {Harmonic plus Yukawa} & {159} & {186} & {203} & {240} \\[1 ex]
		\hline\\[-3.5 ex]
		\multirow{1}{5em}{} & Experimental \cite{patri} & {$203.7(4.7)(0.6)$} & {$-$} & {$257.8(4.1)(0.1)$} & {$-$}\\[1ex]
		\hline\\[-1.3 ex]
		\multirow{11}{10em} {Other theoretical predictions}
		& QCD SR \cite{zhi} & {$208 \pm 10$} & {$263 \pm 21$} & {$240 \pm 10$} & {$308 \pm 21$} \\[1 ex]
		& QCD SR \cite{gel} & {$201^{+12}_{-13}$} &  {$242^{+20}_{-12}$} & {$238^{+13}_{-23}$} & {$293^{+19}_{-14}$}\\[1 ex] 
		& QCD SR \cite{narison} & {$204 \pm 6$} & {$-$} & {$246 \pm 6$} & {$-$} \\[1 ex]
		& LQCD \cite{davies} & {$-$} & {$-$} & {$248 \pm 25$} & {$-$}\\[1 ex] 
		& RQM \cite{capstick} & {$240 \pm 20$} & {$-$} & {$290 \pm 20$} & {$-$} \\[1 ex] 
		& RQM \cite{ebert} & {$234$} & {$310$} & {$268$} & {$315$}\\ [1 ex] 
		& RQM \cite{dae} & {$271 \pm 14$} & {$327 \pm 13$} & {$309 \pm 15$} & {$362 \pm 15$} \\[1 ex] 
		& BS \cite{cevi, guo} &  {$230 \pm 25$} & {$340 \pm 23$} & {$248 \pm 27$} & {$375 \pm 24$} \\[1 ex] 
		& LFQM \cite{chien} & {$-$} & {$259.6 \pm 14.6$} & {$267.4 \pm 17.9$} &{$338.7 \pm 29.7$}\\[1 ex] 
		& LFQM \cite{chao} & {$-$} & {$252^{+13.8}_{-11.6}$} &  {$-$} & {$318.3^{+15.3}_{-12.6}$}\\[1 ex] 
		& Linear\{HO\} \cite{ho} & {$211 \{194\}$} & {$254 \{228\}$} & {$248 \{233\}$} & {$290 \{268\}$}  
		\\[1.5 ex]
		\hline\hline
	\end{tabular}
\end{table}

\end{document}